\documentclass[aps,prb,superscriptaddress,twocolumn,floatfix,notitlepage,a4paper]{revtex4-1}

\usepackage{graphicx}
\usepackage{amsmath,amssymb,amsfonts}
\usepackage{bm}
\usepackage{color}
\usepackage[percent]{overpic}
\usepackage{paralist}
\usepackage[breaklinks=true,colorlinks,citecolor=blue,linkcolor=blue,urlcolor=blue]{hyperref}
\usepackage{comment}

\begin{document}

\title{Electrical plasmon injection in double-layer graphene heterostructures}

\date{\today}

\author{Karina A. Guerrero-Becerra}
\affiliation{Istituto Italiano di Tecnologia, Graphene Labs, Via Morego 30, 16163 Genova, Italy}

\author{Andrea Tomadin}
\affiliation{Department of Physics, Lancaster University, Lancaster LA1 4YB, United Kingdom}

\author{Marco Polini}
\affiliation{Istituto Italiano di Tecnologia, Graphene Labs, Via Morego 30, 16163 Genova, Italy}

\begin{abstract}
It is by now well established that high-quality graphene enables a gate-tunable low-loss plasmonic platform for the efficient confinement, enhancement, and manipulation of optical fields spanning a broad range of frequencies, from the mid infrared to the Terahertz domain.
While all-electrical detection of graphene plasmons has been demonstrated, electrical plasmon injection (EPI), which is crucial to operate nanoplasmonic devices without the encumbrance of a far-field optical apparatus, remains elusive.
In this work, we present a theory of EPI  in double-layer graphene, where a vertical tunnel current excites acoustic and optical plasmon modes.
We first calculate the power delivered by the applied inter-layer voltage bias into these collective modes.
We then
show that this system works also as a spectrally-resolved molecular sensor.
\end{abstract}

\maketitle

\section{Introduction}
\label{sec:introduction}

Recent years have seen rapid progress in the fabrication of van der Waals heterostructures~\cite{geim_nature_2013} comprising graphene, hexagonal boron nitride (hBN), and other two-dimensional (2D) crystals.
These advances have stimulated a large number of theoretical and experimental studies of the optoelectronic properties of these materials and their heterostructures.~\cite{low_natmat_2016, basov_science_2016,koppens_naturenano_2014,ni_nature_2018} A great deal of work has been focused on graphene plasmons, which, in high-quality sheets encapsulated between hBN crystals, have shown truly tantalizing properties.~\cite{low_natmat_2016, basov_science_2016,koppens_naturenano_2014,ni_nature_2018} In view of such properties, it is not hard at all to envision the realization in the near future of a 2D plasmonic platform where plasmon injection, propagation, and detection, occurs in the complete absence of far-field light but is rather achieved via purely electrical methods.

While all-electrical graphene plasmon detection has been recently demonstrated both in the mid infrared~\cite{lundeberg_naturemater_2017} and Terahertz~\cite{pablo_naturenano_2017} spectral ranges, electrical plasmon injection (EPI) remains elusive.
\begin{figure}
\begin{overpic}[width=1.0\linewidth]{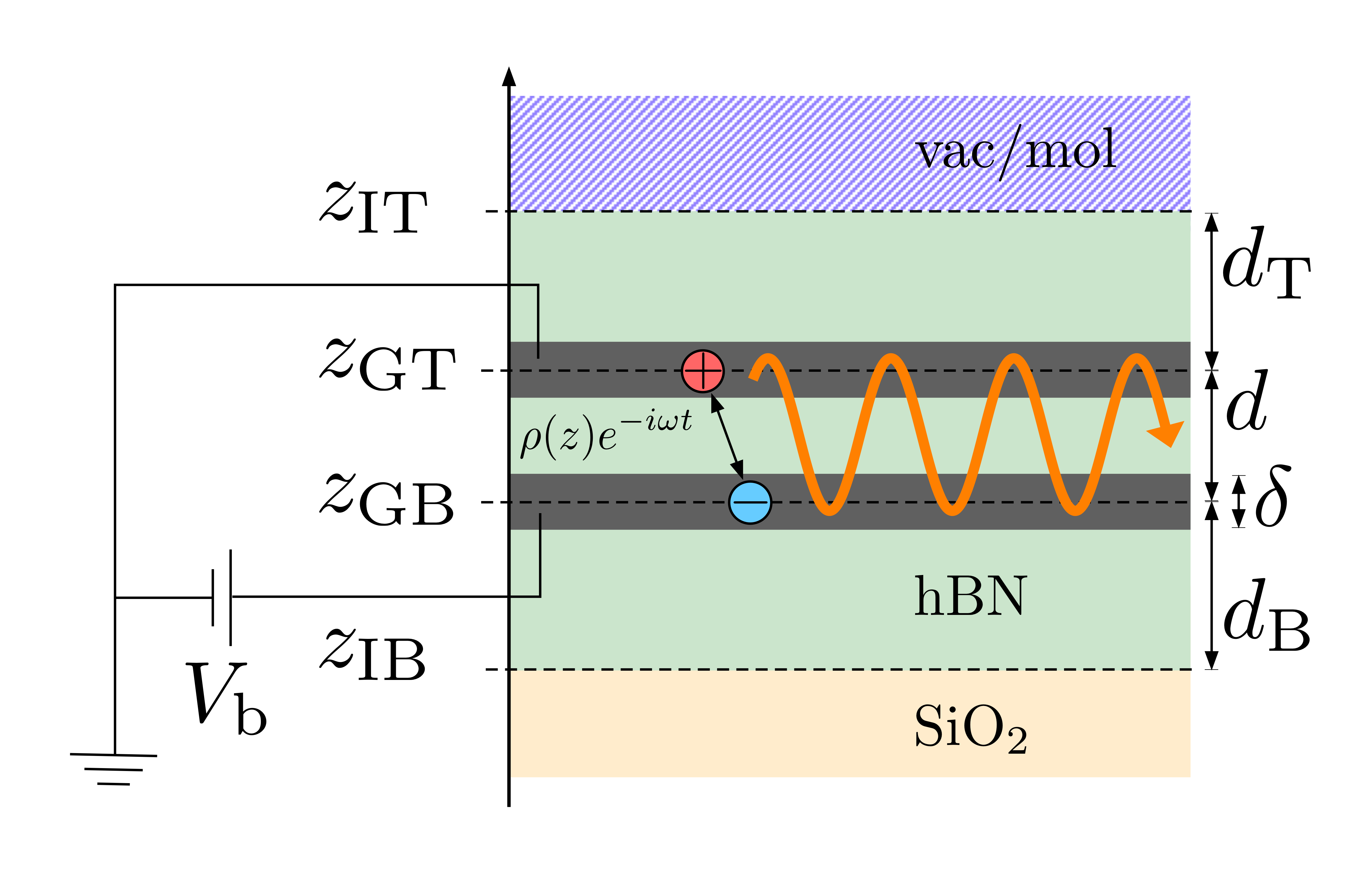}\end{overpic}
\caption{\label{fig:setup}
(Color online) Schematic representation of the double-layer graphene heterostructure studied in this work.
Two graphene sheets lying in the planes $z = z_{\rm GB}$ and $z = z_{\rm GT}$, represented as layers with thickness $\delta$ (gray), are encapsulated by hBN (green) extending from $z_{\rm IB}$ to $z_{\rm IT}$ above a semi-infinite ${\rm SiO}_2$ substrate (yellow).
The thickness of the bottom, middle, and top hBN layer is $d_{\rm B}$, $d$, and $d_{\rm T}$, respectively.
The top semi-space $z > z_{\rm IT}$ is filled with vacuum or a model molecular material (hatched).
An electric bias voltage $V_{\rm b}$ is maintained between the bottom and the top graphene layers by an external source.
Because of the bias voltage, electrons tunnel from the top to the bottom layer, establishing a tunnel current and inducing electric dipolar charges $\rho(z)$ (red and blue circles), oscillating at an angular frequency $\omega$, which couple to the traveling electric field (orange line) of the collective modes of the double-layer heterostructure. }
\end{figure}
A promising route to achieve EPI is offered by a tunnel current between two graphene sheets separated by a thin insulating barrier.
Plasmon emission by tunnel currents has been demonstrated in metal-semiconductor interfaces,~\cite{tsui_prl_1969, tsui_pr_1969} degenerate semiconductors,~\cite{duke_pr_1969} metal-insulator-metal tunnel junctions,~\cite{lambe_prl_1976} and between metallic tips and surfaces.~\cite{berndt_prl_1991}
Early experiments focused on the spectroscopic signatures of plasmon excitations in the tunnel current.~\cite{wolf_rep_prog_phys_1978}
Later on, plasmon excitations were shown~\cite{lambe_prl_1976} to couple the tunnel current to propagating electromagnetic modes, achieving light emission from tunnel junctions.

Recent spectroscopy studies of graphene layers~\cite{brar_prl_2010} and graphene-based heterostructures~\cite{vdovin_prl_2016,ghazaryan_natureelectron_2018} have demonstrated the existence of electron-plasmon interactions, phonon- and magnon-assisted tunneling, respectively.
These studies suggest that the goal of EPI is within reach, and motivate a thorough theoretical investigation of the phenomenon.

In this work, we theoretically study the problem of EPI via a tunnel current.
We consider the double-layer heterostructure depicted in Fig.~\ref{fig:setup}, comprising two graphene sheets encapsulated by hBN, on an ${\rm SiO}_2$ substrate.
The semi-infinite space above the heterostructure consists of either vacuum or a toy-model molecular material with a simple absorption spectrum.
The graphene double-layer supports electronic collective modes~\cite{profumo_prb_2012} (``optical'' and ``acoustic'' plasmons), which hybridize with the optical phonon polaritons of hBN and the molecular excitations.
An external source applies a bias voltage to the two graphene sheets and generates a tunnel current, which feeds the collective modes of the system.
We calculate the power delivered by the tunnel current to the collective modes as the bias voltage is varied, showing that the system works both as a plasmon source and a spectrally-resolved molecular sensor.

The manuscript is organized as follows.
In Sect.~\ref{sec:model} we outline our theoretical formulation, which includes: (i) the quantum mechanical description of the tunneling electrons; (ii) the dielectric functions of the various layers; and (iii) the method to calculate the electric field distribution in the double-layer heterostructure.
In Sect.~\ref{sec:modes} we report analytical expressions for the collective modes of the graphene double-layer, i.e.~the optical and acoustic plasmons, and for the graphene double-layer coupled to a molecular layer, i.e.~a molecular polariton.
The results of our numerical calculations are discussed in Sect.~\ref{sec:results} and compared to the analytical expressions.
Finally, in Sect.~\ref{sec:conclusions} we draw our main conclusions.

\section{Theoretical formulation}
\label{sec:model}

\subsection{Theories of plasmon injection by a tunnel current}
\label{sec:theories}

The calculation of the elastic tunnel current between metallic surfaces separated by a thin insulating layer was first considered by J.~Bardeen in a seminal paper,~\cite{bardeen_prl_1961} which introduced the fundamental concepts that later evolved into the so-called ``transfer-Hamiltonian'' method.~\cite{duke_book_1969}
This method was soon adapted to take into account inelastic tunneling,~\cite{bennet_pr_1968} i.e.~the interaction of tunneling electrons with impurities and collective electronic excitations localized around the insulating layer.
This approach was successfully applied to the case of surface plasmons at metal-semiconductor interfaces as well.~\cite{ngai_prl_1969,economou_prb_1971}

Notwithstanding these early successes, the transfer-Hamiltonian method was the object of several critiques, because a rigorous assessment of its range of validity was missing.~\cite{feuchtwang_book_1978}
The two most criticized points of the theory were the perturbative treatment of the tunneling operator and the precise specification of the ``initial'' and ``final'' single-particle wave functions involved in the tunneling process.
The lack of general agreement on the range of validity of the transfer-Hamiltonian method stimulated several alternative, although related, approaches,~\cite{brailsford_prb_1970,caroli_jpc_1971,feuchtwang_prb_1974} to treat elastic and inelastic tunnel currents.

The theories mentioned above were motivated by experiments using the tunnel current as a spectroscopic tool.~\cite{wolf_rpp_1978}
The existence of plasmon modes (or of other kind of excitations) localized around the insulating layer was taken into account by these theories in the form of inelastic tunneling channels, affecting the tunneling rate and the density of states and, hence, the current-voltage characteristics.
After the experimental demonstration of light-emission from tunnel junctions,~\cite{lambe_prl_1976} however, more work was devoted to the relation between the tunnel current and the intensity of the emitted radiation.
The tunnel current excites plasmon modes at the interfaces, which subsequently couple to propagating electromagnetic modes.
The energy-momentum mismatch between plasmon and propagating modes is overcome if the surfaces are sufficiently rough.
Different plasmon modes at the interface have different roles in this two-step process, coupling more to the tunnel current (``slow'' modes) or to the photonic modes (``fast'' modes).

First, a theory by L.C.~Davis~\cite{davis_prb_1977} explained the light emission from tunnel junctions in terms of the classical coupling between the tunnel current and the electric field of the slow-mode plasmon at the interface.
Then, using the transfer-Hamiltonian method, it was proposed~\cite{hone_apl_1978, rendell_prb_1981} that random fluctuations of the tunnel current drive the slow mode in the insulating layer.
Based on this concept, B.~Laks and D.L.~Mills~\cite{laks_prb_1979, laks_prb_1980} formulated a fruitful theory that allowed, in particular, to discuss the role of the slow and fast plasmon modes in the light-emission process.
Later on, the theory by Laks and Mills was used to study light emission in the more complicated geometry of a scanning tunneling microscope tip in the vicinity of a surface.~\cite{johansson_prb_1990, uehara_jjap_1992}

Very recently, the process of plasmon emission by a tunnel current between graphene sheets was studied in Refs.~\onlinecite{enaldiev_prb_2017,devega_acsphotonics_2017}. Both these works are developed around the concept that tunneling is driven by electron-electron interactions between graphene's carriers.
In Ref.~\onlinecite{enaldiev_prb_2017}, plasmon excitations are encoded in the pole structure of the density-density polarization function of the graphene double layer, which is calculated and related to the tunnel current.
Ref.~\onlinecite{devega_acsphotonics_2017}, instead, uses an effective interaction, obtained as the electric potential produced by an external charge screened by the graphene sheets, treated as conductors with finite conductivity.

In this work, we chose to follow the theoretical approach introduced by Davis,~\cite{davis_prb_1977} which consists of the following steps: (i) calculate the stationary wave function of the tunneling electrons; (ii) derive an electronic charge density, oscillating in a dipolar fashion between the two graphene sheets; and (iii) solve the Poisson equation for the electric field in the heterostructure, using the charge density calculated in the previous step as the source term.
Two features set this method apart from other approaches.~\cite{enaldiev_prb_2017,devega_acsphotonics_2017}
First, there is no notion of an effective interaction between electrons and plasmons, as is common to calculations based on a transfer-Hamiltonian formulation.
The advantage is that our approach circumvents the need of a perturbative expansion in the strength of the light-matter interaction.
Second, the dipolar oscillations of the electronic charge density are purely due to the quantum interference between the stationary wave functions of the tunneling electrons.
In this way, the method separates the calculation of the power delivered to the collective modes from the calculation of the back-action of the electric field on the tunnel current density, which was performed elsewhere.~\cite{enaldiev_prb_2017,devega_acsphotonics_2017}

In the following sections, we provide the elements of the theory which are needed to take the steps (i)--(iii) outlined above.

\subsection{Tunneling-induced dipoles}
\label{sec:tunn_dipoles}

\begin{figure}
\begin{overpic}[width=1.0\linewidth]{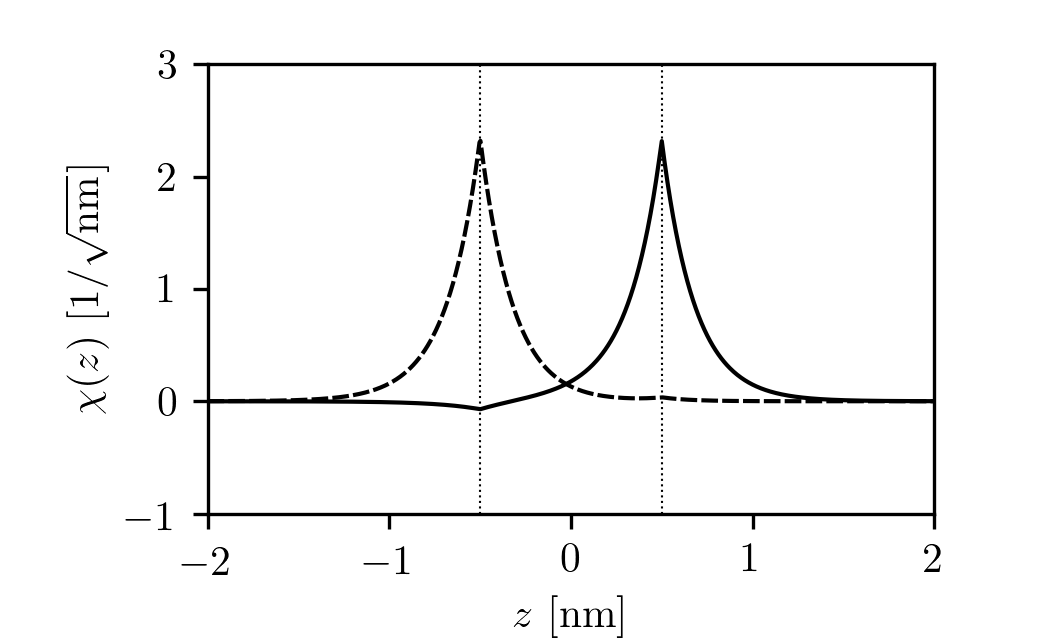}\put(2,55){(a)}\end{overpic}
\begin{overpic}[width=1.0\linewidth]{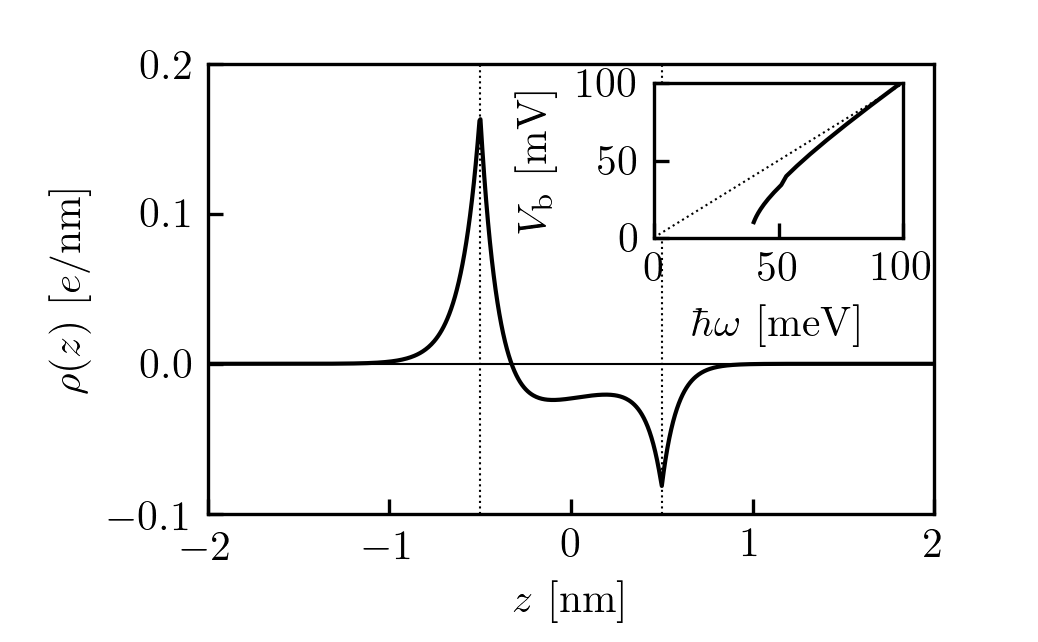}\put(2,55){(b)}\end{overpic}
\caption{\label{fig:charge}
(a) The envelope wave functions $\chi_{\rm i}(z)$ (solid line) and $\chi_{\rm f}(z)$ (dashed line) of the electronic wave functions, for $V_{\rm b} = -1.0~{\rm V}$.
The vertical dotted lines represent the position of the two graphene sheets.
The large bias voltage produces a substantial localization of the two wave functions in the top and bottom graphene sheet, respectively.
(b) The electron charge density $\rho(z)$ defined in Eq.~(\ref{eq:transitioncharge}), which oscillates with angular frequency $\omega$.
The plot demonstrates polarization of opposite charges on the two graphene
sheets, i.e.~an electric dipole along the $z$ direction.
The inset shows the relation between the oscillation energy $\hbar \omega$ and the bias voltage.
The parameters used in the calculation are given in Sect.~\ref{sec:results}. }
\end{figure}

Let us first consider the calculation of the wave function of electrons tunneling in the graphene double-layer.
Along the $z$ direction, the electric potential $U(z)$ experienced by an electron is: (i) constant for $z < z_{\rm GB}$ and $z > z_{\rm GT}$; (ii) linearly varying in the interval $z_{\rm GB} < z < z_{\rm GT}$ because of the bias voltage, ranging from $U(z \to z_{\rm GB}) = -e V_{\rm b} / 2$ to $U(z \to z_{\rm GT}) = e V_{\rm b} / 2$ (where $e$ is the absolute value of the electron charge); and (iii) singular at the positions of the graphene sheets, represented as a negative Dirac delta function with amplitude $2[\hbar^{2} W_{\rm b} / (2 m_{\rm eff})]^{1/2}$, where $m_{\rm eff}$ is the electronic effective mass and $W_{\rm b}$ is the work function of graphene in hBN.
Since the electron wave function is essentially localized around the graphene double-layer, we neglect the interfaces $z = z_{\rm IB}$ and $z = z_{\rm IT}$.
The electron wave function in the heterostructure can be written in the product form
\begin{equation}
\psi({\bm r}, z; t) = \chi(z) e^{i {\bm q} \cdot {\bm r}} e^{- i \varepsilon t / \hbar}~,
\end{equation}
where ${\bm r}$ (${\bm q}$) is a position vector (wave vector) in the graphene plane and the envelope wave function $\chi(z)$ is normalized such that $\int dz |\chi(z)|^{2} = 1$.
The Schr\"odinger equation can be easily solved by separation of variables, using a linear combination of exponentials and Airy functions, and matching boundary conditions at the discontinuities of the potential.
One obtains two bound states $|{\rm i}\rangle$ and $|{\rm f}\rangle$, with envelope wave functions $\chi_{\rm i}(z)$ and $\chi_{\rm f}(z)$, and energies $\varepsilon_{\rm i} > \varepsilon_{\rm f}$.
In this calculation, we neglect the kinetic energy due to the in-plane motion (i.e.~the band dispersion) as well as Fermi statistics.

The crucial step in our approach is to recognize that the electronic system is open, in the sense that it is connected to an external source which injects and extracts electrons.
For this reason, electrons do not persist indefinitely in an eigenstate of the Hamiltonian, but occupy, in general, states which are coherent superpositions of the two eigenstates.
The general electronic wave function then reads
\begin{equation}\label{eq:superposition}
\Psi({\bm r}, z; t) = \alpha_{\rm i} \chi_{\rm i}(z) e^{i {\bm q}_{\rm i} \cdot {\bm r}} e^{-i \varepsilon_{\rm i} t / \hbar} + \alpha_{\rm f} \chi_{\rm f}(z) e^{i {\bm q}_{\rm f}  \cdot  {\bm r}} e^{-i \varepsilon_{\rm f} t / \hbar}~.
\end{equation}
This wave function describes the electronic state until a quantum jump takes place, realizing a tunneling event from the initial $|{\rm i}\rangle$ to the final state $|{\rm f}\rangle$.
Uncorrelated tunneling events build up the total tunnel current between the two graphene layers.
The quantum dissipative dynamics responsible for the quantum jump could be modeled with a quantum master equation,~\cite{gardiner_zoller} with a Lindblad term describing the action of the external source in terms of electron extraction from the upper layer and electron injection into the bottom layer.
In this work, we leave the precise form of the dissipative dynamics unspecified because, for what follows, the values of the coefficients $\alpha_{\rm i, f}$ are not important and it is sufficient to absorb the product $\alpha_{\rm f}^{\ast} \alpha_{\rm i}$ into the definition of an effective density $\bar{n}_{\rm t}$ of tunneling electrons.
We point out that Eq.~(\ref{eq:superposition}) represents a pure state, but the result of the quantum master equation is in general a density matrix $\hat{\rho}$.
In this case, the role of the product  $\alpha_{\rm f}^{\ast} \alpha_{\rm i}$ is played by the ``coherence'' $\langle {\rm f} | \hat{\rho} | {\rm i} \rangle$.
Finally, we notice that, since we neglect the band dispersion, the wave vectors ${\bm q}_{\rm i, f}$ are unrelated.

The charge density derived from the wave function~(\ref{eq:superposition}) is $\rho({\bm r}, z; t) = -e \bar{n}_{\rm t} |\Psi({\bm r}, z; t)|^{2}$.
Upon expanding the squared modulus, one finds two stationary parts, proportional to $|\chi_{\rm i, f}(z)|^{2}$, and two parts oscillating with the transition frequency $\omega = (\varepsilon_{\rm i} - \varepsilon_{\rm f}) / \hbar$ and its complex conjugate.
The oscillating part, which we refer to as ``transition charge density,'' is
\begin{equation}\label{eq:transitioncharge}
\rho_{\rm t}({\bm r}, z; t) = \bar{n}_{\rm t} \rho(z) e^{i {\bm q} \cdot {\bm r}} e^{-i \omega t}~,
\end{equation}
with ${\bm q} = {\bm q}_{\rm i} - {\bm q}_{\rm f}$, and $\rho(z) = - e \chi_{\rm f}(z)^{\ast} \chi_{\rm i}(z)$.
Notice that Eq.~(\ref{eq:transitioncharge}) is in general a complex quantity.
Since the electron wave functions are localized around the position of the graphene planes, the transition charge density has a predominantly dipolar character, as shown in Fig.~\ref{fig:charge}.

In conclusion, electrons, tunneling between bound states, create an electric dipole oscillating at the transition frequency.
The dipole oscillation is of purely quantum origin, because it follows from the superposition in the wave function~(\ref{eq:superposition}).
In the following sections, we study the coupling of the oscillating dipoles to the electric field and the collective modes of the double-layer heterostructure.

\subsection{Dielectric function of the graphene layers}
\label{sec:dielectric_graphene}

In the present and in the following two sections, we detail the dielectric functions of the different layers in the heterostructure of Fig.~\ref{fig:setup}.
To calculate the electric potential, we choose to take into consideration the finite thickness $\delta$ of each graphene sheet, which is smaller than the inter-layer separation $d$, but not negligible.
We then need to provide an effective 3D dielectric function $\epsilon(q, \omega)$ for the finite-thickness graphene layer.
We start out from the random-phase approximation (RPA)~\cite{giuliani_and_vignale_2005}
\begin{equation}\label{eq:rpa_3D}
\epsilon(q, \omega) = 1 - v_{q} \tilde{\chi}^{(0)}(q, \omega)~,
\end{equation}
where $v_{q} = 4 \pi e^{2} / q^{2}$ is the Fourier transform of the Coulomb interaction potential between carriers in the finite-thickness graphene layer and $\tilde{\chi}^{(0)}(q, \omega)$ is the proper non-interacting density-density polarization function (i.e.~the Lindhard function).
To connect this effective Lindhard function to the well-known Lindhard function $\tilde{\chi}_{\rm 2D}^{(0)}(q, \omega)$ of massless Dirac fermions (MDF) in graphene,~\cite{kotov_rmp_2012} we use the linear-response relations~\cite{giuliani_and_vignale_2005} $\bar{n}(q, \omega) = \tilde{\chi}_{\rm 2D}(q, \omega) V_{\rm ext}(q, \omega)$ and $n(q, \omega) = \tilde{\chi}(q, \omega) V_{\rm ext}(q, \omega)$, where $V_{\rm ext}(q, \omega)$ is an external scalar potential and $\bar{n}(q, \omega)$ [$n(q, \omega)$] is the induced density fluctuations in the 2D (finite-thickness) graphene layer.
Neglecting density variations in the $z$ direction, we have $\bar{n}(q, \omega) = \delta \times n(q, \omega)$, which implies $\tilde{\chi}(q, \omega) = \tilde{\chi}_{\rm 2D}(q, \omega) / \delta$.
[The same relation then holds for the non-interacting polarization functions, $\tilde{\chi}^{(0)}(q, \omega) = \tilde{\chi}^{(0)}_{\rm 2D}(q, \omega) / \delta$.]

Substituting these relations into Eq.~(\ref{eq:rpa_3D}), we find
\begin{equation}\label{eq:rpa}
\epsilon(q, \omega) = 1 - \frac{2}{q \delta} v_{{\rm 2D}, q} \tilde{\chi}^{(0)}_{\rm 2D}(q, \omega)~,
\end{equation}
where $v_{{\rm 2D}, q} = 2 \pi e^{2} / q$ is the Fourier transform of the Coulomb interaction potential between MDF in the graphene sheet.
Notice the ``form factor'' $2 / (q \delta)$ which differentiates Eq.~(\ref{eq:rpa}) from the well-known RPA for MDF.
The Lindhard function $\tilde{\chi}^{(0)}_{\rm 2D}$ depends on the average 2D electron density $\bar{n}$.
Here, we assume that the electron density is the same in both graphene sheets and that the Fermi energy lies above the Dirac point.
We have verified that, using Eq.~(\ref{eq:rpa}) in the limit $q \delta \ll 1$, one recovers the correct expressions (see Sect.~\ref{sec:analytical}) for the plasmon spectrum in a single- and double-layer graphene system.

\subsection{Dielectric function of the hBN crystals}

The encapsulant we have chosen for our calculations, hBN, is an anisotropic and uniaxial material, meaning that its dielectric function has different values in the crystal plane $[\epsilon_{xy}(\omega)]$ and in the stacking direction $[\epsilon_{z}(\omega)]$, which are principal directions of the dielectric tensor.~\cite{dai_science_2014,caldwell_naturecommun_2014}
Moreover, hBN is a natural hyperbolic material, i.e.~there are frequency ranges, called \emph{reststrahlen} bands, where $\epsilon_{xy}(\omega)$ and $\epsilon_{z}(\omega)$ have different signs, producing a peculiar propagation of the electric field.~\cite{dai_science_2014,caldwell_naturecommun_2014}
These properties are captured by the following dielectric function
\begin{equation}\label{eq:eps_hbn}
\epsilon_{\alpha}(\omega) = \epsilon_{\alpha,\infty} + \frac{(\epsilon_{\alpha,0} - \epsilon_{\alpha,\infty})\omega_{\alpha,{\rm T}}^{2}}{\omega_{\alpha,{\rm T}}^{2} - \omega^{2} - i \gamma_{\alpha} \omega}~,
\end{equation}
where $\alpha = xy$ or $\alpha = z$.
Here, $\epsilon_{\alpha,\infty}$ and $\epsilon_{\alpha,0}$ are high- and low-frequency dielectric constants, $\omega_{\alpha, {\rm T}}$ is the frequency of transverse optical phonon-polaritons in the $\alpha$ direction, and $\gamma_{\alpha}$ is the corresponding damping rate.
The frequencies of the corresponding longitudinal modes are given by the Lyddane-Sachs-Teller relation~\cite{grosso_book} $\omega_{\alpha, {\rm L}} = \omega_{\alpha, {\rm T}} [\epsilon_{\alpha,0} / \epsilon_{\alpha,\infty}]^{1/2} > \omega_{\alpha, {\rm T}}$.
The lower (upper) reststrahlen band is located in the frequency range $\omega_{z, {\rm T}} < \omega < \omega_{z, {\rm L}}$ ($\omega_{xy, {\rm T}} < \omega < \omega_{xy, {\rm L}}$).

\subsection{Dielectric function of the molecular layer}
\label{sec:analytical}

For the dielectric function of the molecular material at $z > z_{\rm IT}$ we use the expression
\begin{equation}\label{eq:eps_mol}
\epsilon_{\rm mol}(\omega) = \epsilon_{\infty} + \frac{(\epsilon_{0} - \epsilon_{\infty}) \Omega_{0}^{2}}{\Omega_{0}^{2} - \omega^{2} - i \gamma \omega / \hbar}~.
\end{equation}
This expression, similar in form to Eq.~(\ref{eq:eps_hbn}), is easily derived starting from the equation of motion of the position vector ${\bm R}$ of a bound electron in the presence of an electric field ${\bm E}$,
\begin{equation}
\ddot{\bm R}(t) = (-e) {\bm E} / m - \Omega_{0}^{2} {\bm R}(t) - \gamma \dot{\bm R}(t) / \hbar~,
\end{equation}
where $\hbar \Omega_{0}$ is the energy of the electronic resonance, $\gamma$ is its broadening, $m$ is an effective electronic mass, and all the variables oscillate with angular frequency $\omega$.
From the expression for the steady-state polarization ${\bm P}_{\rm mol} = -e n_{\rm mol} {\bm R}$, with $n_{\rm mol}$ the three-dimensional molecular density, one derives the polarization $\chi_{\rm mol}(\omega)$ such that ${\bm P}_{\rm mol} = \chi_{\rm mol}(\omega) {\bm E}$, and then the dielectric function $\epsilon_{\rm mol}(\omega) = \epsilon_{\infty} + 4 \pi \chi_{\rm mol}(\omega)$.
We neglect local-field effects such as those described by the well-known Clausius-Mossotti formula.~\cite{grosso_book}
The high-frequency dielectric constant $\epsilon_{\infty}$ encodes the small-scale details of the molecular material while the low-frequency constant $\epsilon_{0} = \epsilon_{\rm mol}(0)$ is identified from Eq.~(\ref{eq:eps_mol}) as $\epsilon_{0} = \epsilon_{\infty} + 4 \pi e^{2} n_{\rm mol} / (m 	\Omega_{0}^{2}) > \epsilon_{\infty}$.
In practice, the dielectric function is more easily specified by treating $\Omega_{0}$, $\gamma$, $\epsilon_{0}$, and $\epsilon_{\infty}$ as independent constants.

\subsection{Electric potential in the double-layer heterostructure}
\label{sec:potential}

The electric potential $\phi({\bm r}, z; t) = \phi(z) e^{-i {\bm q} \cdot {\bm r}} e^{-i \omega t}$ in the double-layer heterostructure, in the presence of a tunnel current, is obtained by solving the Poisson equation
\begin{equation}
- \nabla \cdot \left [ \epsilon_{\ell}(\omega) \nabla \phi({\bm r}, z; t) \right ] = 4 \pi \rho_{\rm t}({\bm r}, z; t)~,
\end{equation}
with the transition charge density~(\ref{eq:transitioncharge}) as the source.
The dielectric function $\epsilon_{\ell}(\omega)$ of each layer $\ell$ has been described in the previous sections.

Let us summarize the method to solve the Poisson equation in the heterostructure,~\cite{maier_book_2007} which amounts to a transfer-matrix approach accounting for the presence of the source.
In each layer $z_{\ell} < z < z_{\ell + 1}$ the electric potential is written as $\phi_{\ell}(z) = \alpha_{\ell} e^{-q_{\ell}(z - z_{\ell})} + \beta_{\ell} e^{-q_{\ell}(z_{\ell + 1} - z)} + g_{\ell}(z)$, where the function $g_{\ell}(z)$ solves the Poisson equation in the $\ell$th layer without taking into account the boundary conditions.
The boundary conditions state that: (i) the electric potential is continuous at the interfaces; (ii) the field vanishes away from the heterostructure; and (iii) the $z$ component of the displacement field is continuous at the interfaces.
The last boundary condition holds because, in the Davis' approach,~\cite{davis_prb_1977} the conducting regions (in our case, the graphene layers) are effectively treated as dielectrics, i.e.~the electronic polarization is taken into account in the dielectric function~(\ref{eq:rpa}) and does not generate free charges at the interfaces of the heterostructure.
Applying the boundary conditions, we first find that $q_{\ell} = q [\epsilon_{xy}(\omega)/\epsilon_{z}(\omega)]^{1/2}$ in the hBN layers and $q_{\ell} = q$ otherwise, and we then obtain a linear system $L({\bm q}, \omega)$ of $12$ equations for $12$ unknowns $\{\alpha_{\ell}$, $\beta_{\ell}\}_{\ell = 1}^{6}$, for each pair of values ${\bm q}$, $\omega$, which we solve numerically.

A solution of the linear system $L({\bm q}, \omega)$ is found for any wave vector ${\bm q}$ and frequency $\omega$ because the Poisson equation yields the field $\phi({\bm r}, z; t)$ produced by a given charge density $\rho_{\rm t}({\bm r}, z; t)$.
On the other hand, to find the collective modes of the heterostructure, i.e.~the self-sustained oscillations of the electric field, one has to solve the Laplace equation, i.e.~the Poisson equation with charge density set to zero.
At fixed ${\bm q}$, the Laplace equation can be solved only for a discrete set of values of $\omega$, corresponding to the angular frequencies of the collective modes.
Finding the solutions of the Laplace equation at fixed ${\bm q}$ is thus analogous to calculating the eigenvalues and eigenvectors of a secular equation, with the added considerable difficulty that here the dependence on the eigenvalues (i.e.~$\omega$) is nonlinear.
In practice, we proceed by numerically finding the roots of the function of $\omega$ defined as follows:
(i) we set $\beta_{6} = 1$;
(ii) we solve the reduced linear system given by the first $11$ equations;
(iii) we calculate the value of the $12$th equation.
When this function is zero, then all $12$ equations are solved and $\omega$ is an eigenfrequency of the system.
In this procedure, the ordering of the equations and variables is arbitrary, however, we find a higher numerical accuracy by including in the reduced linear system the equations which represent the continuity of the potential at the interfaces.

Finally, the power $P$ per area $A$ delivered by the tunnel current to the collective modes reads
\begin{equation}\label{eq:dissipation}
\frac{P}{A} = 2 \omega \bar{n}_{\rm t} \int_{z_{\rm IB}}^{z_{\rm IT}} dz {\rm Im}[\phi(z)^{\ast} \rho(z)]~.
\end{equation}
It could be surprising that in Eq.~(\ref{eq:dissipation}) the absorption is due to the same charge density $\rho(z)$ which generates the potential $\phi(z)$.
A more careful look, however, shows that the phase between $\phi(z)$ and $\rho(z)$, which makes the integral non-vanishing, is due to the imaginary part of the dielectric functions.
In other words, Eq.~(\ref{eq:dissipation}) represents the energy dissipated into the electronic and molecular degrees of freedom of the heterostructure.
Since we solve the Poisson equation, ignoring retardation in the Maxwell equations, the electric field that we calculate does not describe coupling to the far-field modes, so the contribution of radiation losses is not present in Eq.~(\ref{eq:dissipation}).

\section{Collective modes of the double-layer heterostructure}
\label{sec:modes}
A graphene double-layer, in a uniform medium with dielectric constant $\bar{\epsilon}$, supports a high-energy ``optical'' plasmon mode with dispersion~\cite{profumo_prb_2012}
\begin{equation}\label{eq:optical_plasmon}
\hbar \omega_{\rm op}(q\to 0) =  \sqrt{N_{\rm f} \varepsilon_{\rm F} e^{2} q / (\sqrt{2} \bar{\epsilon})}~,
\end{equation}
where $N_{\rm f} = 4$ is number of fermion flavors and $\varepsilon_{\rm F}$ is the Fermi energy.
The latter is given by $\varepsilon_{\rm F}= \hbar v_{\rm F} k_{\rm F}$, with the Fermi wave vector $k_{\rm F} = \sqrt{\pi \bar{n}}$ and Fermi velocity $v_{\rm F}$.~\cite{katsnelson_book}
(We reiterate our assumption that the electron density is the same in both graphene sheets and that the Fermi energy lies above the Dirac point.)
This mode corresponds to the plasmon mode of a single layer with twice the density.
The double-layer graphene supports also a low-energy acoustic mode with dispersion~\cite{profumo_prb_2012}
\begin{equation}\label{eq:acoustic_plasmon}
\hbar \omega_{\rm ac}(q \to 0) =  \hbar v_{\rm F} q \frac{1 + k_{\rm TF} d}{\sqrt{1 + 2 k_{\rm TF} d}}~,
\end{equation}
where $k_{\rm TF} = 4 k_{\rm F} \alpha_{\rm ee}$ is the Thomas-Fermi wave vector, with $\alpha_{\rm ee} = e^{2} / (\bar{\epsilon} \hbar v_{\rm F})$ the electron-electron coupling strength.
Making contact to the jargon used in works concerned with light emission in tunnel junctions (cfr.~Sect.~\ref{sec:theories}), the optical and acoustic mode are the ``fast'' and ``slow'' mode of the heterostructure, respectively.

In the presence of a semi-space characterized by the dielectric function~(\ref{eq:eps_mol}), the optical and acoustic mode hybridize with the molecular oscillations.
It is easy to see that a new collective ``polariton'' mode appears in the spectrum, with a dispersion that tends to
\begin{equation}\label{eq:molecular_polariton}
\hbar \omega_{\rm mp}(q) \to \hbar \Omega_{0} \sqrt{\frac{\epsilon_{0} + \bar{\epsilon}}{\epsilon_{\infty} + \bar{\epsilon}}}~.
\end{equation}
The previous expression turns out to be valid \emph{both} in the short-wavelength $q d$, $q d_{\rm T} \gg 1$ and long-wavelength $q d$, $q d_{\rm T} \ll 1$ limits.

The analysis of the collective modes of the double-layer heterostructure and of the delivered power is complicated by the hyperbolic nature of the hBN.
Indeed, a thick hBN slab acts as a Fabry-Perot resonator, where the electric potential oscillates between the interfaces with an arbitrary large number of nodes.~\cite{tomadin_prl_2015}
All these modes accumulate towards the lower (upper) extreme of the upper (lower) reststrahlen band and separate as the wave vector $q$ increases.
Some of these modes strongly hybridize with the plasmon modes supported by the graphene layers and the polariton mode supported by the molecular oscillations.
In Eqs.~(\ref{eq:optical_plasmon}), (\ref{eq:acoustic_plasmon}), and~(\ref{eq:molecular_polariton}) we have used a uniform, frequency-independent dielectric constant $\bar{\epsilon}$.
To use those formulas in the presence of the hBN layers, and gain a qualitative analytical understanding of the collective modes, one needs to take $\bar{\epsilon} = \sqrt{\epsilon_{xy,0} \epsilon_{z,0}}$.

\begin{figure}
\begin{overpic}[width=1.0\linewidth]{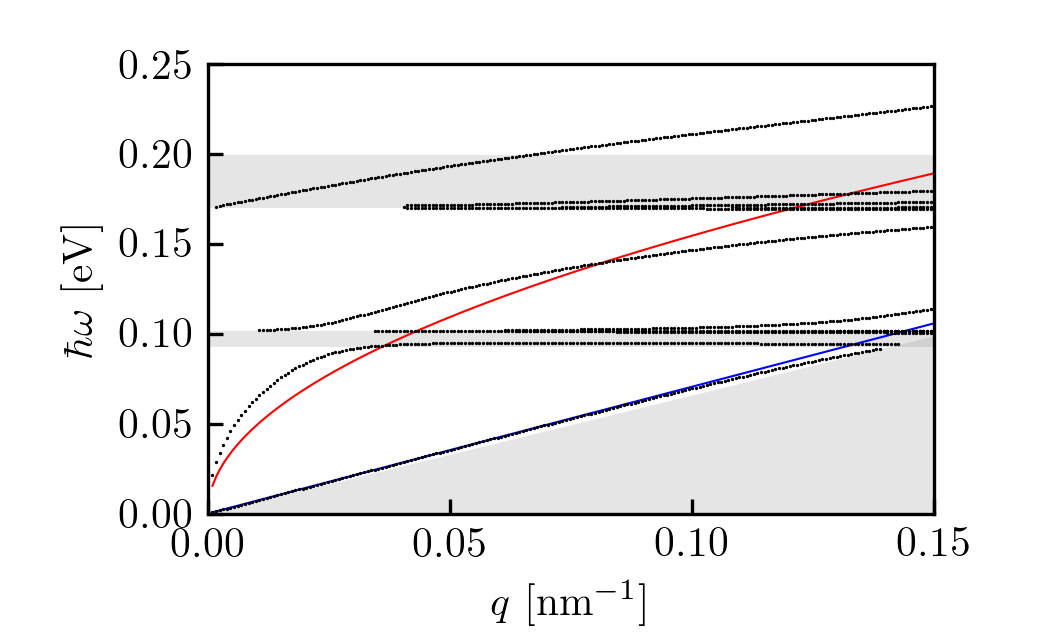}\put(2,55){(a)}\end{overpic}
\begin{overpic}[width=1.0\linewidth]{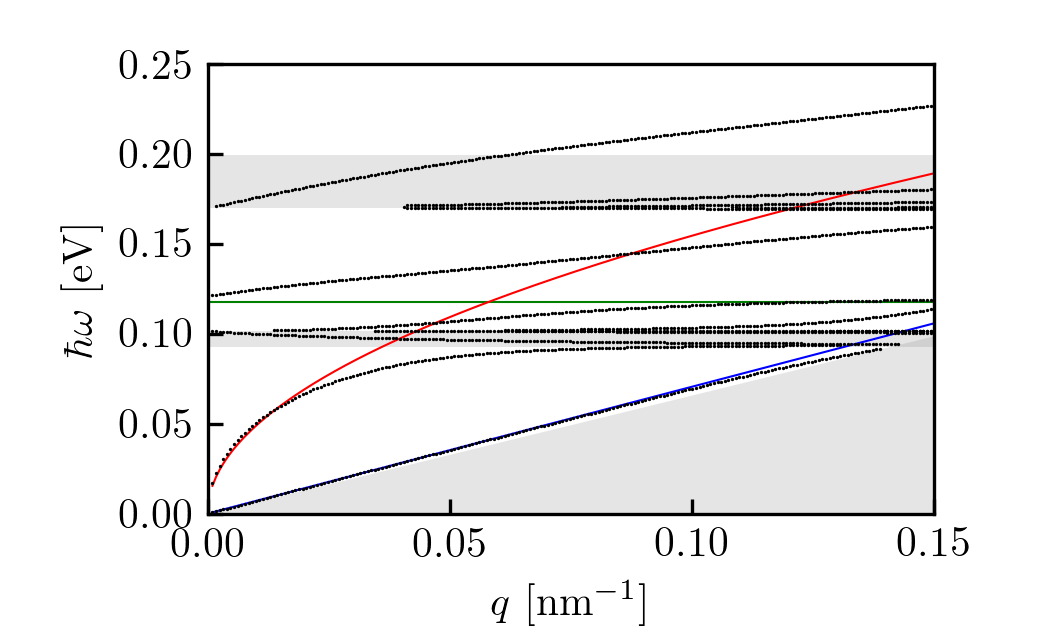}\put(2,55){(b)}\end{overpic}
\caption{\label{fig:dispersion}
(Color online) Dispersion of the collective modes of the double-layer heterostructure (points), when the top semi-space is filled with vacuum (a) or the model molecular material (b) defined in Sect.~\ref{sec:model}.
Gray-shaded areas correspond, from low to high values of $\hbar \omega$, to the intra-band electron-hole continuum,~\cite{kotov_rmp_2012} the lower and the upper reststrahlen bands of the hBN layers.
The red, blue, and green solid lines correspond to the analytical dispersion of the optical plasmon Eq.~(\ref{eq:optical_plasmon}), acoustic plasmon Eq.~(\ref{eq:acoustic_plasmon}), and molecular polariton Eq.~(\ref{eq:molecular_polariton}), respectively. }
\end{figure}

\begin{figure}
\begin{overpic}[width=0.49\linewidth]{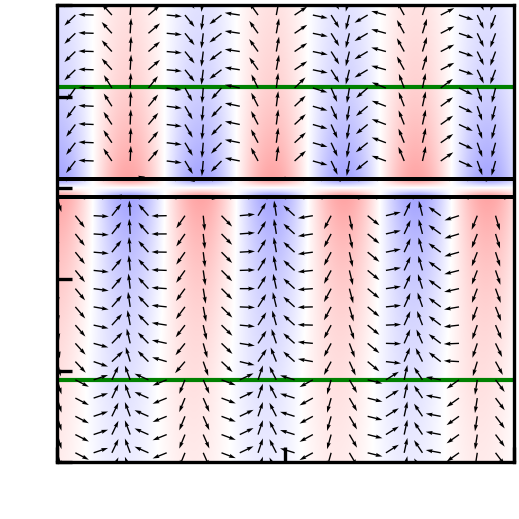}\put(0,95){(a)}\put(30,3){$\hbar \omega = 70~{\rm meV}$}\end{overpic}
\begin{overpic}[width=0.49\linewidth]{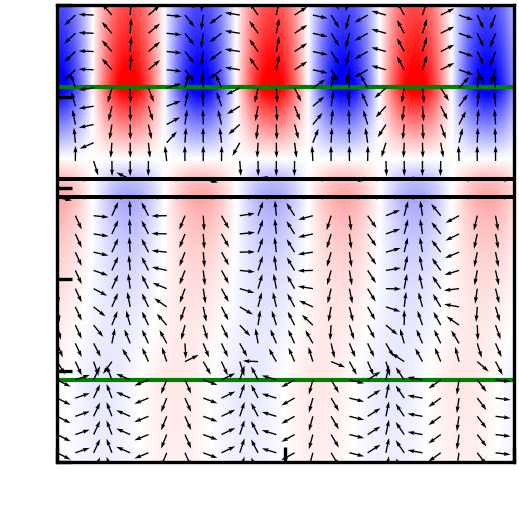}\put(0,95){(b)}\put(30,3){$\hbar \omega = 116~{\rm meV}$}\end{overpic}
\begin{overpic}[width=0.49\linewidth]{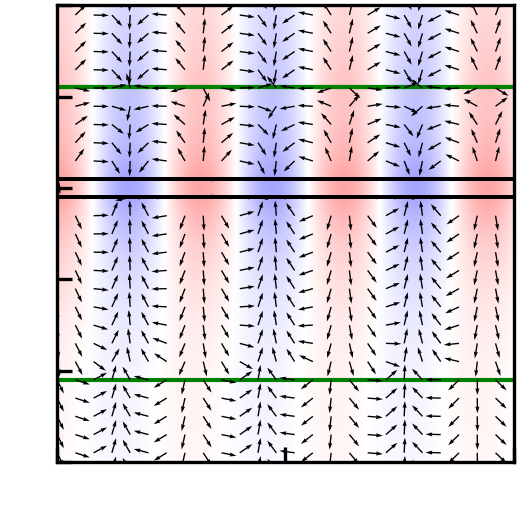}\put(0,95){(c)}\put(30,3){$\hbar \omega = 148~{\rm meV}$}\end{overpic}
\begin{overpic}[width=0.49\linewidth]{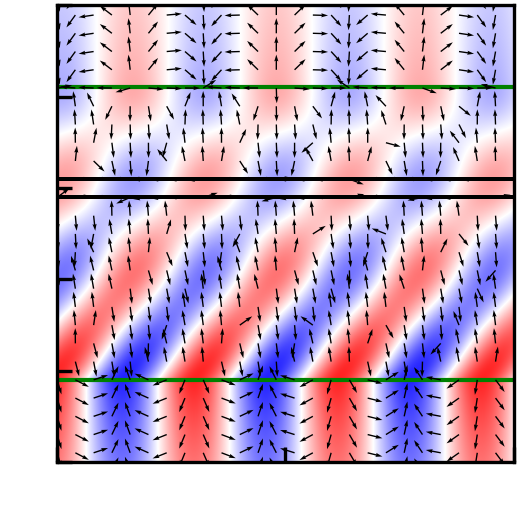}\put(0,95){(d)}\put(30,3){$\hbar \omega = 171~{\rm meV}$}\end{overpic}
\caption{\label{fig:field}
(Color online) Space profile of the electric potential (color map) and of the direction of the electric field (arrows) as a function of $x$ (horizontal axis) and $z$ (vertical axis), in the range $0 < x <200~{\rm nm}$ and $-15~{\rm nm} < z < 10~{\rm nm}$.
For graphical convenience, the axis labels are not shown.
Red (blue) shades correspond to positive (negative) electric potential.
The horizontal black lines denote the locations $z_{\rm GB}$, $z_{\rm GT}$ of the graphene sheets and the horizontal green lines the interfaces $z_{\rm IB}$, $z_{\rm IT}$ of hBN.
The fields correspond to the modes shown in Fig.~\ref{fig:dispersion}(b) at $q = 0.1~{\rm nm}^{-1}$, and at the energy indicated at the bottom of each panel.
(a)-(c) show three modes outside of the reststrahlen bands, corresponding to the acoustic plasmon, molecular polariton, and the optical plasmon, respectively.
(d) Shows a mode within the upper reststrahlen band, exhibiting Fabry-Perot oscillations~\cite{tomadin_prl_2015} in the hBN layers. }
\end{figure}

\begin{figure}
\begin{overpic}[width=1.0\linewidth]{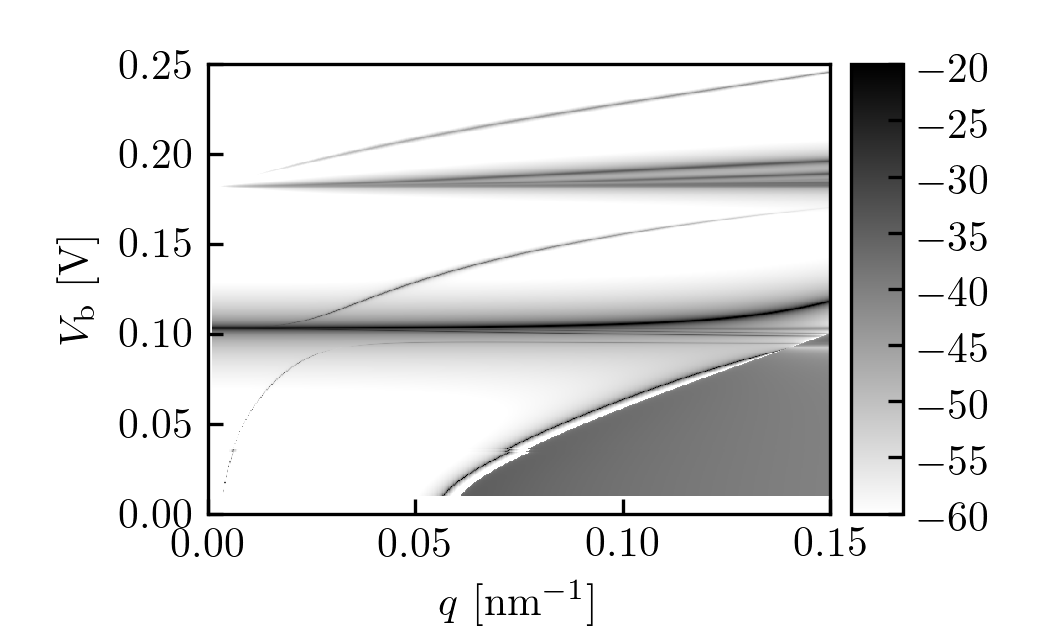}\put(2,55){(a)}\end{overpic}
\begin{overpic}[width=1.0\linewidth]{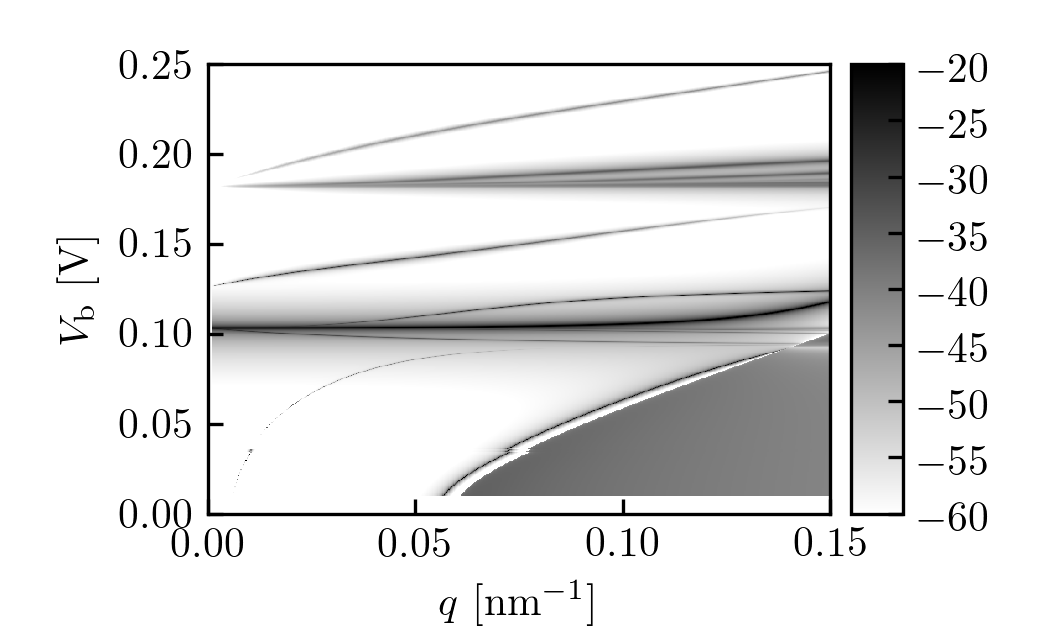}\put(2,55){(b)}\end{overpic}
\caption{\label{fig:emission}
Power per unit area $P/A$ delivered by the tunnel current to the collective modes of the double-layer heterostructure with wave vector $q$ (horizontal axis), as the bias voltage $V_{\rm b}$ is tuned (vertical axis), when the top semi-space is filled with vacuum (a) or the model molecular material (b) defined in Sect.~\ref{sec:model}.
The colorbar represents $P/A$ divided by its maximum in ${\rm dB}$.
The intra-band electron-hole continuum and the reststrahlen bands of the hBN layers are clearly visible as extended regions of high absorption.
These region appear distorted with respect to the gray-shaded ares in Fig.~\ref{fig:dispersion} because the relation between $\hbar \omega$ and $V_{\rm b}$ is not linear [see inset of Fig.~\ref{fig:charge}(b)].
Between these regions, sharp continuous features are easily identified with the collective modes shown in Fig.~\ref{fig:dispersion}.
The minor discontinuity around $V_{\rm b} \simeq 0.4~{\rm V}$ is due to the numerical implementation of Airy functions at large arguments. }
\end{figure}

\begin{figure}
\begin{overpic}[width=1.0\linewidth]{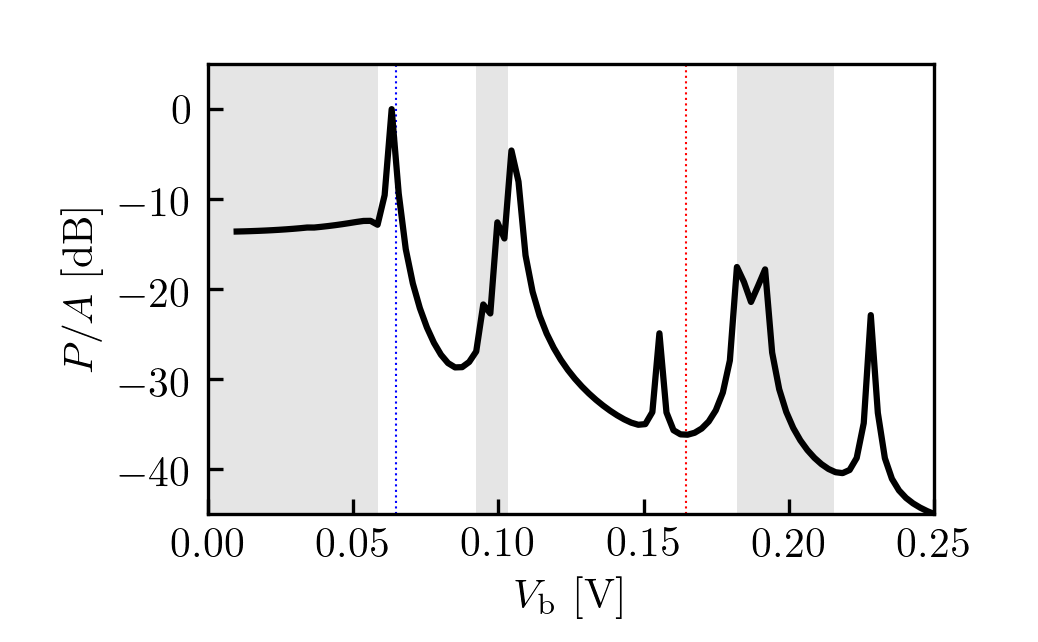}\put(2,55){(a)}\end{overpic}
\begin{overpic}[width=1.0\linewidth]{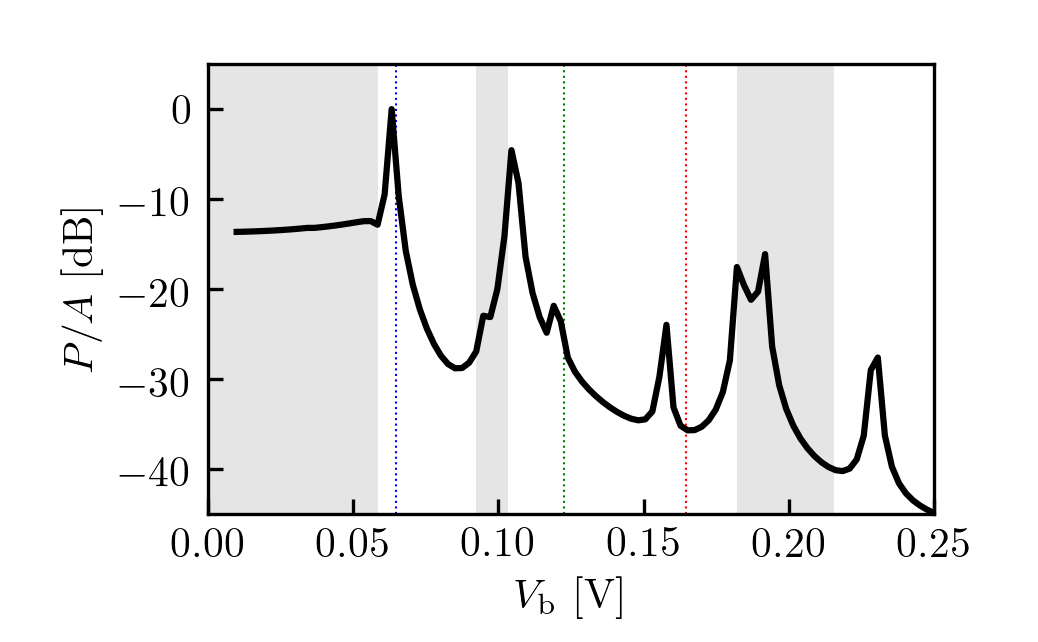}\put(2,55){(b)}\end{overpic}
\caption{\label{fig:emission_cuts}
(Color online) Power per unit area $P/A$ delivered by the tunnel current to the collective modes of the double-layer heterostructure with wave vector $q = 0.1~{\rm nm}^{-1}$ as the electric bias voltage $V_{\rm b}$ is changed, when the top semi-space is filled with vacuum (a) or the model molecular material (b) defined in Sect.~\ref{sec:model}.
Gray-shaded areas correspond, from low to high values of $V_{\rm b}$, to the intra-band electron-hole continuum, and the lower and upper reststrahlen bands of the hBN layers.
Vertical dotted lines correspond to the value of the bias voltage where a peak of the absorption due to a collective mode is expected, according to the simplified analytical formulas discussed in Sect.~\ref{sec:modes}.
These modes are, from low to high values of $V_{\rm b}$, the acoustic plasmon Eq.~(\ref{eq:acoustic_plasmon}), the molecular polariton Eq.~(\ref{eq:molecular_polariton}), and the optical plasmon Eq.~(\ref{eq:optical_plasmon}). }
\end{figure}

\section{Numerical results}
\label{sec:results}

We  now turn to illustrate the main results obtained by numerically solving the model outlined in Sect.~\ref{sec:model}.
Our goal is to show that the peaks of the absorption spectrum, i.e.~the magnitude of $P/A$ given in Eq.~(\ref{eq:dissipation}) as a function of the bias voltage $V_{\rm b}$, correspond to collective modes of the double-layer heterostructure.

For convenience, we summarize in this paragraph all the parameters that we use in the calculation.
The geometry of the double-layer heterostructure is defined by
$d = 1.0~{\rm nm}$,
$d_{\rm B} = 10.0~{\rm nm}$, and
$d_{\rm T} = 5.0~{\rm nm}$.
For the electron density in the graphene sheets we take
$\bar{n} = 3.0\times 10^{12}~{\rm cm}^{-2}$.
The parameters of the negative Dirac delta function potential at the position of the graphene sheets, introduced in Sec.~\ref{sec:tunn_dipoles}, are chosen $W_{\rm b} = 2.25~{\rm eV}$~\cite{kharche_nanolet_2011} (assuming that the bands of graphene and hBN are aligned~\cite{}) and $m_{\rm eff} = 0.5~m_{\rm e}$.~\cite{britnell_science_2012}
The finite thickness of the graphene sheets, introduced in Sec.~\ref{sec:dielectric_graphene}, is taken to be $\delta = 0.2~{\rm nm}$.
The dielectric constant of the substrate is
$\epsilon_{{\rm Si}{\rm O}_2} = 3.9$;
for the molecular ensemble we take the reasonable values
$\epsilon_{0} = 4.0$,
$\epsilon_{\infty} = 1.5$,
$\hbar \Omega_{0} = 100.0~{\rm meV}$, and
$\gamma = 0$;
and for the hBN layers we use~\cite{cai_solidstatecomm_2007}
$\epsilon_{x,\infty} = 4.87$,
$\epsilon_{z,\infty} = 2.95$,
$\epsilon_{x,0} = 6.70$,
$\epsilon_{z,0} = 3.56$,
$\gamma_{xy} = 0.87~{\rm meV}$,
$\gamma_{z} = 0.25~{\rm meV}$,
$\hbar \omega_{z,{\rm T}} = 92.5~{\rm meV}$,
$\hbar \omega_{z,{\rm L}} = 101.6~{\rm meV}$,
$\hbar \omega_{xy,{\rm T}} = 170.1~{\rm meV}$, and
$\hbar \omega_{xy,{\rm L}} = 199.5~{\rm meV}$.

Fig.~\ref{fig:dispersion} shows the dispersion of the collective modes, calculated on a mesh of wave vectors as explained in Sect.~\ref{sec:potential}.
The rich structure of Fabry-Perot-like modes in the reststrahlen bands is prominent.
However, outside of the reststrahlen bands and the intra-band electron-hole continuum, the optical and acoustic plasmon and the molecular polariton are clearly identifiable.
The acoustic plasmon is less hybridized with other modes, being very close to the graphene intra-band continuum, and the analytical expression Eq.~(\ref{eq:acoustic_plasmon}) proves accurate in the whole displayed interval.
For the optical plasmon, the expression in Eq.~(\ref{eq:optical_plasmon}) gives a very good approximation of the numerical result in the long-wavelength limit.
Between the reststrahlen bands, however, where the dispersion of the hybridized mode is much flattened, the analytical expression crosses the numerical results in the neighborhood of $q \simeq 0.1~{\rm nm}^{-1}$ (for the parameters used here).
The expression in Eq.~(\ref{eq:molecular_polariton}) for the molecular polariton correctly captures the long-wavelength limit of the hybrid mode which, for larger wave vectors, becomes the optical plasmon between the reststrahlen bands.
A different mode splits off from the lower reststrahlen band and converges to the molecular polariton expression for $q \gtrsim 0.05~{\rm nm}^{-1}$.

The nature of the modes is demonstrated in Fig.~\ref{fig:field}, where the space profile of the electric potential and the direction of the electric field is shown at fixed wave vector.
For the sake of the graphical representation, the length of the arrows is not proportional to the magnitude of the electric field.
The acoustic plasmon [panel (a)] is characterized by electric potential of opposite sign on the top and bottom graphene sheets.
Between the sheets, the field is thus mostly directed along $z$.
The force lines of the field are almost unperturbed at the interface with the molecular material.
For this reason, hybridization between the acoustic plasmon and the molecular polariton is absent.
For the optical plasmon [panel (c)], the electric potential has the same sign on the top and bottom graphene sheets.
The field is thus mostly directed in the $x-y$ plane between the two graphene sheets.
The behavior of the force lines of the field is different at the interface $z = z_{\rm IB}$ with the substrate ${\rm Si}{\rm O}_{2}$ and $z = z_{\rm IT}$ with the molecular material.
Indeed, as Fig.~\ref{fig:dispersion} shows, the hybridization between the optical plasmon and the molecular polariton is strong.
The molecular polariton mode [panel (b)] is easily identified because the electric potential is strongest at the interface $z = z_{\rm IT}$.
For completeness, we also show a typical Fabry-Perot-like mode within a reststrahlen band [panel (d)].
The mode is characterized by periodic oscillations of the electric potential along $z$, which appear as diagonal stripes of constant potential.
The profile of the potential is only slightly perturbed by the presence of the double-layer, so that the mode resonates between the interfaces $z = z_{\rm IB}$ and $z = z_{\rm IT}$, in the whole region occupied by hBN.

Fig.~\ref{fig:emission} shows the absorption spectrum on a wave vectors mesh.
It is important to notice that, since one cannot span the entire $\hbar \omega$ range by tuning $V_{\rm b}$ [see inset of Fig.~\ref{fig:charge}(b)], the vertical axes in Figs.~\ref{fig:dispersion} and~\ref{fig:emission} are not linearly proportional.
However, a one-to-one correspondence between the peaks of the absorption spectrum and the collective modes can be easily drawn.
This figure clearly shows that, by driving a tunnel current between the graphene sheets, one excites the collective modes of the double-layer heterostructure.
Moreover, the nearby presence of a molecular layer changes the absorption spectrum, which means that the system acts as a frequency-resolved plasmon-enabled detector.~\cite{maier_book_2007,brolo_naturephoton_2012}
These are the main results of this work.

Fig.~\ref{fig:emission_cuts} shows the absorption spectrum at fixed wave vector $q = 0.1~{\rm nm}^{-1}$, i.e.~vertical cuts from Fig.~\ref{fig:emission}, normalized to its maximum value.
The peaks corresponding to absorption by the acoustic plasmon, molecular polariton, and optical plasmon are identified by comparison with the analytical expression which, as shown in Fig.~\ref{fig:dispersion}, are sufficiently accurate in this wave vector range.
We see that the largest absorption is associated with the acoustic plasmon.
This feature can be understood by inspecting the space profile of the electric fields in Fig.~\ref{fig:field}.
The electric field of the acoustic plasmon is largest between the graphene sheets and directed along $z$, and thus it is optimally coupled to the oscillating dipoles generated by the tunneling electrons [see~Fig.~\ref{fig:charge}(b)].
This observation is in agreement with the results of Refs.~\onlinecite{laks_prb_1979, laks_prb_1980}, if one remembers that the acoustic mode is the ``slow'' mode of the graphene-based heterostructure.
Notwithstanding the dominance of the acoustic-plasmon peak, the peak corresponding to the molecular polariton between the reststrahlen bands is also clearly visible.
Finally, since the electron density along $z$ is not purely anti-symmetric [see Fig.~\ref{fig:charge}(b)], due to the finite bias which breaks space inversion around $z = 0$, the optical plasmon mode can also be excited.

\section{Conclusions}
\label{sec:conclusions}

In conclusion, in this work we have calculated the absorption spectrum of a double-layer graphene heterostructure, where a tunnel current between the graphene layers is generated by an external source.
In our theoretical approach, the tunneling electrons generate oscillating dipoles which couple to the collective modes of the double-layer heterostructure, i.e.~the acoustic and optical plasmons and a molecular polariton mode.
This approach highlights the purely quantum nature of the charge density oscillations coupling to the electric field of the plasmon and polariton modes.
We have verified that the peaks of the absorption spectrum correspond to the collective modes of the heterostructure.

Our results show that the setup that we consider can be used \emph{both} as a plasmon source and as a frequency-resolved plasmon-enabled detector.~\cite{maier_book_2007,brolo_naturephoton_2012}
In the first case, we find that acoustic plasmons absorb more power than the other modes due to a better spatial coupling between the field and the oscillating dipoles, and hence are more likely to be excited.
In the second case, the position and resonance frequency of a nearby molecular layer manifests itself as a distinct peak in the absorption spectrum -- see Fig.~\ref{fig:emission_cuts}.
In this case, coupling between the tunnel current and the molecular layer is mediated mostly by the optical plasmon mode, whose field extends further away from the graphene double layer, as shown in Fig.~\ref{fig:field}.
The detection of the molecular resonant frequency is possible in a very large band where the optical plasmon is not overdamped, as long as it does not fall in the reststrahlen bands of hBN, where larger absorption takes place.
However, we reckon that, in the reststrahlen bands,~\cite{dai_science_2014,caldwell_naturecommun_2014} one could use the hyperlensing phenomenon~\cite{li_natcommun_2015,dai_natcommmun_2015,giles_natmater_2018} to couple the optical plasmon to subwavelength absorbers, or to guide the field of the generated modes in a preferred direction.

\acknowledgments

This work was supported by the European Union's Horizon 2020 research and innovation programme under grant agreement No. 785219 - ``GrapheneCore2''. Fruitful discussions with Silvia Dante, Frank Koppens, Kostya Novoselov, and Dmitry Svintsov are gratefully acknowledged.

\end{document}